\documentstyle[12pt,epsfig]{article}
\def\bq{\begin{quote}} 
\def\eq{\end{quote}}

\def\bq{\begin{quote}}

\def\p{\pi}

\catcode`\@=11
\def\@normalsize{\@setsize\normalsize{15pt}\xiipt\@xiipt
\abovedisplayskip 14pt plus3pt minus3pt%

\belowdisplayskip \abovedisplayskip
\abovedisplayshortskip  \z@ plus3pt%
\belowdisplayshortskip  7pt plus3.5pt minus0pt}
\def\small{\@setsize\small{13.6pt}\xipt\@xipt
\abovedisplayskip 13pt plus3pt minus3pt%
\belowdisplayskip \abovedisplayskip
\abovedisplayshortskip  \z@ plus3pt%
\belowdisplayshortskip  7pt plus3.5pt minus0pt
\def\@listi{\parsep 4.5pt plus 2pt minus 1pt
            \itemsep \parsep
            \topsep 9pt plus 3pt minus 3pt}}
 
\def\underline#1{\relax\ifmmode\@@underline#1\else
        $\@@underline{\hbox{#1}}$\relax\fi}
\@twosidetrue
\relax

\catcode`@=12

\catcode`\@=11

\def\section{\@startsection{section}{1}{\z@}{3.5ex plus 1ex minus
   .2ex}{2.3ex plus .2ex}{\large\bf}}

\def\ps@headings{\def\@oddfoot{}\def\@evenfoot{}
\def\@oddhead{\hbox{}\hfill
        \makebox[.5\textwidth]{\raggedright\ignorespaces --\thepage{}--
        \hfill }}
\def\@evenhead{\@oddhead}
\def\subsectionmark##1{\markboth{##1}{}}
}
 
\ps@headings
 
\catcode`\@=12
 
\relax
{
\def\figcap{\section*{Figure Captions\markboth
        {FIGURECAPTIONS}{FIGURECAPTIONS}}\list
        {Fig. \arabic{enumi}:\hfill}{\settowidth\labelwidth{Fig. 999:}
        \leftmargin\labelwidth
        \advance\leftmargin\labelsep\usecounter{enumi}}}
 \relax
\def\tablecap{\section*{Table Captions\markboth
        {TABLECAPTIONS}{TABLECAPTIONS}}\list
        {Table \arabic{enumi}:\hfill}{\settowidth\labelwidth{Table 999:}
        \leftmargin\labelwidth
        \advance\leftmargin\labelsep\usecounter{enumi}}}
 \relax
\def\reflist{\section*{References\markboth
        {REFLIST}{REFLIST}}\list
        {[\arabic{enumi}]\hfill}{\settowidth\labelwidth{[999]}
        \leftmargin\labelwidth
        \advance\leftmargin\labelsep\usecounter{enumi}}}
 \relax
 
\catcode`\@=11
 
\def\marginnote#1{}
\newcount\hour
\newcount\minute
\newtoks\amorpm
\hour=\time\divide\hour by60
\minute=\time{\multiply\hour by60 \global\advance\minute by-
\hour}
\edef\standardtime{{\ifnum\hour<12 \global\amorpm={am}%
    \else\global\amorpm={pm}\advance\hour by-12 \fi
    \ifnum\hour=0 \hour=12 \fi
    \number\hour:\ifnum\minute<100\fi\number\minute\the\amorpm}}
\edef\militarytime{\number\hour:\ifnum\minute<100\fi\number\minute}

\def\draftlabel#1{{\@bsphack\if@filesw {\let\thepage\relax
  \xdef\@gtempa{\write\@auxout{\string
    \newlabel{#1}{{\@currentlabel}{\thepage}}}}}\@gtempa
    \if@nobreak \ifvmode\nobreak\fi\fi\fi\@esphack}
     \gdef\@eqnlabel{#1}}
\def\@eqnlabel{}
\def\@vacuum{}
\def\draftmarginnote#1{\marginpar{\raggedright\scriptsize\tt#1}}
\def\draft{\oddsidemargin -.5truein
        \def\@oddfoot{\sl preliminary draft \hfil
        \rm\thepage\hfil\sl\today\quad\militarytime}
        \let\@evenfoot\@oddfoot \overfullrule 3pt
        \let\label=\draftlabel
        \let\marginnote=\draftmarginnote
 
\def\@eqnnum{(\theequation)\rlap{\kern\marginparsep\tt\@eqnlabel}%
\global\let\@eqnlabel\@vacuum}  }
\def\preprint{\twocolumn\sloppy\flushbottom\parindent 1em
        \leftmargini 2em\leftmarginv .5em\leftmarginvi .5em
        \oddsidemargin -.5in    \evensidemargin -.5in
        \columnsep 15mm \footheight 0pt
        \textwidth 250mmin      \topmargin  -.4in
        \headheight 12pt \topskip .4in
        \textheight 175mm
        \footskip 0pt
 
\def\@oddhead{\thepage\hfil\addtocounter{page}{1}\thepage}
        \let\@evenhead\@oddhead \def\@oddfoot{} \def\@evenfoot{}
}
\def\titlepage{\@restonecolfalse\if@twocolumn\@restonecoltrue\onecolumn
     \else \newpage \fi \thispagestyle{empty}\c@page\z@
        \def\thefootnote{\fnsymbol{footnote}} }
\def\endtitlepage{\if@restonecol\twocolumn \else  \fi
        \def\thefootnote{\arabic{footnote}}
        \setcounter{footnote}{0}}  
\catcode`@=12
\relax
 
\def\ps@headings{\def\@oddfoot{}\def\@evenfoot{}
\def\@oddhead{\hbox{}\hfill
        \makebox[.5\textwidth]{\raggedright\ignorespaces --\thepage{}--
        \hfill }}
\def\@evenhead{\@oddhead}
\def\subsectionmark##1{\markboth{##1}{}}
}
 
\ps@headings
 
\relax

\def\p{\pi}

\def\bo{{\raise.15ex\hbox{\large$\Box$}}}               
\def\face{{\raise.2ex\hbox{$\displaystyle \bigodot$}\mskip-2.2mu \llap {$\ddot
        \smile$}}}                                      

\def\leftrightarrowfill{$\mathsurround=0pt \mathord\leftarrow \mkern-6mu
        \cleaders\hbox{$\mkern-2mu \mathord- \mkern-2mu$}\hfill
        \mkern-6mu \mathord\rightarrow$}       
\def\dvec#1{\vbox{\ialign{##\crcr
        \leftrightarrowfill\crcr\noalign{\kern-1pt\nointerlineskip}
        $\hfil\displaystyle{#1}\hfil$\crcr}}}           

\def\beqx{\begin{displaymath}}
\def\eeqx{\end{displaymath}}

\newcommand{\newc}{\newcommand}
\newc{\ra}{\rightarrow}
\newc{\lra}{\leftrightarrow}
\newc{\beq}{\begin{equation}}
\newc{\eeq}{\end{equation}}
\newc{\bea}{\begin{eqnarray}}
\newc{\eea}{\end{eqnarray}}

\newc{\sm}{Standard Model}
\newc{\smd}{Standard Model}
\newc{\barr}{\begin{eqnarray}}
 \newc{\earr}{\end{eqnarray}}

\begin{document}

\begin{flushright}
hep-ph/9904279 \\  
CERN-TH/99-87 \\
\end{flushright}
\vspace{15mm}

\begin{center}
{\LARGE \bf 
Can Neutrinos be Degenerate in Mass?
} \\
\end{center}

\vspace*{0.7cm}

\begin{center}
{\large {\bf John Ellis} \footnote{
{\tt email: John.Ellis@cern.ch}}
and {\bf Smaragda Lola} \footnote{
{\tt email: magda@mail.cern.ch}} }
\end{center}

\vspace*{0.1cm}
\begin{center}
{\large  Theory Division, CERN, CH 1211 Geneva 23, Switzerland}
\end{center}

\vspace{1 cm}

\begin{center}
{\bf ABSTRACT}
\end{center}

{{\small
We reconsider the possibility that the masses of the three light neutrinos
of the Standard Model might be almost degenerate and close to the
present upper limits from Tritium $\beta$ decay and cosmology. In such
a scenario, the cancellations required by the latest upper limit on
neutrinoless double-$\beta$ decay enforce near-maximal mixing that may
be compatible only with the vacuum-oscillation scenario for solar
neutrinos. We argue that the mixing angles yielded by
degenerate neutrino mass-matrix textures are not in general stable
under small perturbations. We evaluate within the MSSM the 
generation-dependent one-loop renormalization of
neutrino mass-matrix textures
that yielded degenerate masses and large mixing at the tree level.
We find that 
$m_{\nu_e} > m_{\nu_\mu} > m_{\nu_\tau}$
after renormalization, excluding MSW effects
on solar neutrinos. We verify that bimaximal
mixing is not stable, and show that the renormalized masses and mixing
angles are not compatible with all the experimental constraints,
even for $\tan \beta$ as low as unity. These results hold whether 
 the neutrino masses are generated by a see-saw
mechanism with heavy neutrinos weighing $\sim 10^{13}$~GeV 
or by
non-renormalizable interactions at a scale $\sim 10^5$~GeV.
We also comment on the corresponding renormalization effects in the
minimal Standard Model, in which
$m_{\nu_e} < m_{\nu_\mu} < m_{\nu_\tau}$. Although 
a solar MSW effect
is now possible, the perturbed neutrino masses and mixings
are still not compatible with atmospheric- and solar-neutrino data.}}

\noindent

\thispagestyle{empty}

\setcounter{page}{0}
\vfill\eject

\section{Introduction}

Observations of solar and atmospheric neutrinos provide
strong indications that neutrinos may oscillate between
eigenstates with different masses $m_{\nu_i}$
\cite{SKam,KamMac}. The indications from
solar neutrinos are for a difference in mass squared
$\Delta m^2_{solar} \sim 10^{-4}$~\cite{MSW} to
$10^{-10}$~eV$^2$ \cite{solar}, whereas the atmospheric neutrino data
favour $\Delta m^2_{atmo} \sim 10^{-2}$ to $10^{-3}$~eV$^2$ \cite{SKam}.
As is well known, oscillation experiments are not able to set the
overall scale of the neutrino masses. However, there are some
upper limits on these: astrophysical and cosmological
constraints on dark matter suggest that $\Sigma_i m_{\nu_i}
<$ few~eV \cite{Lyalpha},
and experiments on the endpoint of the Tritium $\beta$-decay
spectrum suggest that $m_{\nu_i} < 2.5$~eV~\cite{Lobashev} for any mass
eigenstate
with a substantial electron flavour component~\cite{Barger}. 
The question then
arises whether there are any indirect arguments bearing on
the possibility that the three light neutrino masses might be
approximately degenerate \cite{ol}-\cite{nonab}
and close to these upper limits.

Extending previous arguments in~\cite{GG} to include a new
upper limit on neutrinoless double-$\beta$ decay~\cite{KKG},
we argue in the next section that, within the context of
such degenerate neutrinos, this may already exclude
the large-angle Mikheyev-Smirnov-Wolfenstein
(MSW)~\cite{MSW} solution to the solar-neutrino problem.
In this case, one would be forced into the vacuum-oscillation
solution, and hence extreme mass degeneracy to one part in $10^{10}$.
In the case of the large-angle MSW solution, the degeneracy would
need to be to one part in $10^4$ or more.

We subsequently discuss general features of the
one-loop renormalization of neutrino mass matrices, using as an
example one specific degenerate mass-matrix texture
that accommodates neutrinoless double-$\beta$ decay via bimaximal mixing
\cite{new1,GG}.
We argue that
this and other degenerate textures are generically unstable with
respect to small perturbations, so that mixing does not remain
bimaximal as suggested by
oscillation data and the neutrinoless double-$\beta$ decay constraint.
We evaluate within the Minimal Supersymmetric extension of the
Standard Model (MSSM) the one-loop renormalization-group
corrections to mass degeneracy in scenarios where the masses
are generated either by a see-saw mechanism \cite{seesaw}
at some high scale $M_N$
close to $M_{GUT}$, or by an effective operator at some low scale 
$\Lambda$ close
to $M_{SUSY}$. We find that these corrections are
significant, even for relatively small values of
$\tan\beta$, and lead to the ordering
$m_{\nu_e} > m_{\nu_\mu} > m_{\nu_\tau}$.
In the case that neutrino
masses of order ${\cal O} (2)$~eV arise
via the see-saw mechanism, these corrections
indicate that degenerate neutrinos are not
compatible with all constraints, even for 
tan$\beta$ as low as 1. In particular, we note that
the ordering of neutrino masses is {\it incompatible}
with MSW solutions, because they require
$m_{\nu_e} < m_{\nu_\mu}, m_{\nu_\tau}$.
If neutrino masses arise
from non-renormalizable interactions at a scale 
$\Lambda$  as low as 10-100 $m_{SUSY}$,
the effects are smaller. Still, even in this case,
we cannot obtain the required degeneracy for
the vacuum oscillations, and
MSW oscillations cannot be obtained.

Finally, we discuss briefly renormalization effects
in the  minimal Standard Model. Their expected
magnitude is similar to that in the MSSM in the low-$\tan\beta$ regime,
since for a single Higgs field all the quark and charged-lepton mass
hierarchies have to arise purely from the
Yukawa couplings, and hence the $\tau$ coupling
is small. However, in this case the 
sign of the Yukawa renormalization effects is
opposite, tending to
increase  the magnitudes of the entries in $m_{eff}$
and leading to $m_{\nu_e} < m_{\nu_\mu}, m_{\nu_\tau}$,
as required by the MSW mechanism. However, bimaximal mixing
again cannot be maintained, so that this framework also appears
incompatible with the combined   constraints from 
neutrinoless double-$\beta$ decay and neutrino oscillation
data.

This analysis shows that schemes with degenerate
neutrinos are very problematic, contrary to solutions with
large neutrino hierarchies. Since the former
may be obtained from non-Abelian symmetry structures
\cite{nonab} and the
latter from Abelian ones~\cite{abel,LLR,Grah}, renormalization-group 
effects on the neutrino mass eigenvalues may be providing important
information about the underlying flavour structure of the
fundamental theory.

\section{Neutrinoless Double-$\beta$ Constraints on Degenerate Neutrinos
\label{sec:neutr}}

It was pointed out by Georgi and Glashow~\cite{GG}
that, if the neutrino masses are close to the Tritium and
cosmological upper limit so that relic neutrinos
contribute at least 
one percent of the critical density of the Universe, then
the upper limits on neutrinoless double-$\beta$ decay require the
mixing angle for solar neutrino oscillations to be
almost maximal. This is because the neutrinoless double-$\beta$ decay
limit
constrains the $ee$ component of the Majorana neutrino mass matrix in the
charged-lepton flavour basis~\cite{GG}:
\bea
m_{eff}^{ee} 
\equiv
\big\vert  m_1\,c_2^2c_3^2e^{i\phi} + m_2\,c_2^2s_3^2e^{i\phi'} + m_3\,
s_2^2\,e^{i2\delta} \big\vert < B \,.
\label{doublebeta}
\eea
where $c_i,s_i$ denote $\cos\theta_{i},\sin\theta_{i}$ in the
conventional $3 \times 3$ mixing parametrization
\bea
\pmatrix{\nu_e \cr \nu_\mu\cr \nu_\tau\cr}=
\pmatrix{c_2c_3 &   c_2s_3 &   s_2e^{-i\delta}\cr
-c_1s_3-s_1s_2c_3e^{i\delta} &  
+c_1c_3-s_1s_2s_3e^{i\delta} &  s_1c_2\cr
+s_1s_3-c_1s_2c_3e^{i\delta} &  
-s_1c_3-c_1s_2s_3e^{i\delta} &   c_1c_2\cr}\,
\pmatrix{\nu_1\cr \nu_2\cr \nu_3\cr}\,,
\label{param}
\eea
where the diagonal matrix $m_{eff}^{diag}$ is 
$diag(m_1e^{i\phi}, m_2e^{i\phi'}, m_3)$,
$\phi$ and
$\phi'$ are phases in the light Majorana mass matrix, and
$B$ is the experimental upper
bound on $m_{eff}^{ee}$.
In schemes with degenerate neutrinos, the
differences between the mass eigenvalues $m_i$ in (\ref{doublebeta}) may
be neglected~\cite{GG}. Moreover,
given the upper limits on atmospheric oscillations
into electron neutrinos established by Chooz 
\cite{chooz} and
Super-Kamiokande \cite{SKam}
, we follow~\cite{GG} and set $\theta_2 \approx 0$.
Thus (\ref{doublebeta}) may be simplified to the form:
\bea
\big\vert \cos^2{\theta_3}\,e^{i\phi} + \sin^2{\theta_3}\,e^{i\phi'}
\big\vert < {B \over {\overline m} }
\label{simple}
\eea
where ${\overline m} \approx 2$~eV is the 
conjectured common mass scale
of the (almost-)degenerate neutrinos.

At the time of~\cite{GG}, the best available upper limit was
$B < 0.46$~eV, and the constraint (\ref{simple})
could be satisfied for $\phi+\phi'\simeq \pi$
and $\vert\cos{2\theta_3}\vert < 0.23$, leading to
$\sin^2{2\theta_3}> 0.95$. This was incompatible with
the small-angle MSW solution for the solar neutrino data,
but consistent with either the large-angle MSW solution
or the vacuum-oscillation solution.

Recently, however, a new upper limit $B < 0.2$~eV
has been given~\cite{KKG}, from which we infer
$\vert\cos{2\theta_3}\vert < 0.1$ and hence
\beq
\sin^2{2\theta_3}> 0.99.
\label{KKG}
\eeq
Such maximal mixing is favoured by the vacuum-oscillation
solution, but is disfavoured in the large-angle
MSW solution, because $\sin^2{2\theta_3} = 1$ would yield
an energy-independent suppression of all solar
neutrinos~\cite{Giunti}. This disagrees with the Homestake data
by at least three standard deviations, although the question
persists whether the data could tolerate the lower
limit in (\ref{KKG}). Several global fits to the
solar-neutrino data have been published, including
one by the Super-Kamiokande collaboration~\cite{SKdn} that takes
into account its recent measurements of the day-night
effect and yields $\sin^2{2\theta_3} < 0.99$~\footnote{If
the day-night effect were not included, the upper limit
would be $\sin^2{2\theta_3} < 0.95$: the magnitude of the
day-night effect improves the quality of the fit in this
large-angle MSW region~\cite{SKdn}.}. Other global fits often give
smaller upper limits on $\sin^2{2\theta_3}$ 
\cite{SKdn}.

We infer, provisionally, that the large-angle MSW
solution is now excluded by neutrinoless double-$\beta$ decay
if the neutrinos are near-degenerate,
forcing us into the vacuum-oscillation solution,
in which case the neutrino mass degeneracy must be at the
level of one part in $10^{10}$. We could, however,
imagine possible ways to evade this conclusion.
Perhaps a small but non-trivial admixture of $\nu_{\mu} - \nu_e$
atmospheric oscillations could soften the bound
(\ref{KKG}), and/or perhaps the solar-neutrino data could be
stretched to accommodate it: $\sin^2{2\theta_3} = 1$ is excluded
just at the $99.8 \%$ confidence level. Alternatively, the
constraint (\ref{KKG}) would be weakened for degenerate
neutrinos weighing less than 2~eV. We comment later how
our conclusions would be affected if the large-angle MSW
solution could be tolerated, in which case the neutrino mass
degeneracy need only be to one part in $10^4$.

Using the experimental information then available,
a specific effective neutrino mass texture was 
proposed in~\cite{new1,GG}:
\bea
{m_{eff}} \propto \,\pmatrix{
0 &  {1\over\sqrt2} &  {1\over\sqrt2}\cr
{1\over\sqrt2} &  {1\over2} &  -{1\over2}\cr
{1\over\sqrt2} &  -{1\over2} &  {1\over2}\cr
}
\label{GGtexture}
\eea
in the flavour basis where charged-lepton masses are diagonal,
leading to bimaximal mixing.
Other neutrino mass textures might be considered,
depending on the theoretical assumptions and other 
phenomenological choices made.
In the following, we shall use (\ref{GGtexture}) as an
example, but frame our discussion in terms sufficiently
general that it could be extended to other model textures.

Any such texture can only be regarded as a first approximation,
that might be modified by higher-order effects. These could
include the possible contributions of 
higher-dimensional non-renormalizable operators. The above
discussion suggests that any such contributions should 
change the mass eigenstates by at most one part in $10^{10}$
(or $10^4$),
which is considerably more delicate than the expected hierarchy
$m_{GUT}/m_{P} \approx 10^{-2}$. In the absence of any detailed
theory of such contributions, one cannot say that this is 
necessarily a problem. However, global symmetries are not
normally expected to be exact at the Planck scale, so the
mass-degeneracy constraint is potentially powerful.

Moreover, the mixing angles in such degenerate mass-matrix
models are inherently unstable when higher-order
perturbations are switched on, as we discuss in more detail later.

\section{Renormalization-Group Effects on Neutrino Mass Textures}

Calculable and potentially significant breakings of
the neutrino mass degeneracies
are provided by
renormalization-group effects. However, these
depend on the specific neutrino model framework
adopted. We consider here two possibilities: one
is the conventional see-saw, with a singlet-neutrino
mass scale $M_N \sim 10^{13}$~GeV
\footnote{
We note that the heavy
Majorana neutrino masses $M_N$
need not be degenerate.
On the contrary, flavour symmetries indicate
that should have a structure determined by
the flavour charges of the $N$ fields, and
of other singlet fields in a given model.
However, we do not discuss explicitly here this
structure, which could also in principle affect the amount of
renormalization-group running.}, and the other is
a model where the light Majorana neutrino masses are
simply generated by a new non-renormalizable interaction,
such as $\nu_L \nu_L H H$,
at a mass scale $\Lambda \sim 10^5$~GeV,
close to $m_{SUSY} = 10^3$~GeV. Below we
give numerical results for both scenarios.

In the see-saw case,
between the GUT scale and the scale $M_{N}$ of the
heavy Majorana neutrinos,
there is an effect on the mixing angle
due to the renormalization-group running  of the
Dirac neutrino coupling $Y_N$ \cite{VB}:
 \begin{eqnarray}
  8\p^2 {d\over dt}({ Y}_N { Y}_N^\dagger)& = &
 \{ -\sum_i c_N^i g_i^2+3 ({ Y}_N { Y}_N^\dagger)
+{\rm Tr}[3({ Y}_U { Y}_U^\dagger)+({ Y}_N { Y}_N^\dagger)] \}
({ Y}_N { Y}_N^\dagger)  \nonumber \\
       & &\qquad \qquad   +{1\over 2}\{({ Y}_E { Y}_E^\dagger)
 ({ Y}_N { Y}_N^\dagger)+({ Y}_N { Y}_N^\dagger)
({ Y}_E { Y}_E^\dagger)\}  
\label{firstRG} 
\end{eqnarray}
Here and subsequently we work at the one-loop level, and denote
the renormalization-group scale by $t \equiv \ln{\mu}$.
In the MSSM, $c^i_N = (3/5, 3, 0)$,
and we denote the Dirac couplings of other types of fermion $F$ by $Y_F$.
It is apparent from (\ref{firstRG}) that
large Yukawa couplings 
have a bigger effect on 
$m^{D}_{33}$  than on the rest of the mass-matrix
elements, and tend in general to lower $Y_N$.
This alters the structure of the Dirac mass matrix,
in turn affecting the magnitudes
of the mixing angles. These effects
become more relevant in examples where
cancellations between various entries 
may lead to amplified mixing in $m_{eff}$.
However, we assume here that any neutrino-mass texture~\cite{GG}
is defined at the characteristic scale $M_N$ of the see-saw
mechanism. Therefore, in the current
discussion we use this first part of the run only in order
to define the initial conditions for the gauge and Yukawa
couplings at $M_N$, but not to modify the neutrino-mass texture.
Since the exact form of $m_{eff}$ depends on
the right-handed Majorana mass matrix, we 
simply assume that this has the form that is required
in order to lead to a specific texture at
$M_N$.

Having set the initial conditions for
$g_i$ and $\lambda_i$ at $M_N$, we note that
${ Y}_N$ decouples 
below the right-handed Majorana-mass scale, 
where the relevant running
is that of the effective neutrino-mass operator
\cite{Bab,run}:
 \begin{equation} 
    8\p^2 {d\over dt}{ m_{eff}} = \{-({3\over 5} g_1^2+3 g_2^2)+
{\rm Tr} [3{ Y}_U { Y}_U^\dagger]\}m_{eff}
 +{1\over 2}\{({ Y}_E { Y}_E^\dagger)m_{eff} +
   m_{eff} ({ Y}_E { Y}_E^\dagger)^T\}       \ ,
\label{MASSES}
\end{equation}
This is the basic equation for the running
of the various entries of the effective light-neutrino mass matrix.
We continue to work in the basis where the 
charged-lepton mass matrix is diagonal,
which is also the basis in which the neutrino-mass texture
is specified.

In previous work \cite{ELLN2}, we discussed the running of
the (23) mixing angle, but here we focus more
on the evolution of the $m_{eff}$ entries themselves,
extending our previous discussion to encompass
solutions to the solar neutrino problem. 
For this purpose, we use the following differential
equations for individual elements
of  the effective neutrino-mass matrix:
\bea
\frac{1}{m_{eff}^{ij}}\frac{d}{d t}m_{eff}^{ij}&=& 
\frac{1}{8\pi^2}\left( -c_i g_i^2 + 3 h_t^2 
+ \frac{1}{2} ( h_{i}^2 + h_j^2)   \right) 
\label{RGmeff}
\eea
It is convenient for the subsequent discussion to
define the integrals
\bea
 I_g& =& exp[\frac{1}{8\pi^2}\int_{t_0}^t(-c_i g_i^2 dt)]\\
I_t &=& exp[\frac{1}{8\pi^2}\int_{t_0}^t  h_t^2 dt]\\
    I_{h_i} &=& exp[\frac 1{8\pi^2}\int_{t_0}^t h_{i}^2 dt]
\eea
where in $I_{h_i}$ and $h_i$
the subindex $i$ refers to the 
charged-lepton flavours $e, \mu$ and $\tau$.
Simple integration of (\ref{RGmeff}) yields
\bea
\frac{m_{eff}^{ij}}{m_{eff,0}^{ij}}&=& 
 exp\left\{ \frac{1}{8\pi^2}\int_{t_0}^t
             \left(-c_i g_i^2 + 3 h_t^2 
+ \frac{1}{2} ( h_{i}^2 + h_j^2)
\right)\right\}
\nonumber\\
    &=& I_g\cdot I_t \cdot 
 \sqrt{I_{h_i}} \cdot  \sqrt{I_{h_j}}
\label{RGis}
\eea
where the initial conditions are denoted by $ m_{eff,0}^{ij}$.
As we have already noted, these conditions are defined at
$M_N$, the scale where the neutrino Dirac coupling $h_N$ decouples from
the renormalisation-group equations. 

Using (\ref{RGis}), we see that an initial texture $m_{eff,0}^{ij}$
at $M_N$ is modified to become
\bea
m_{eff}  = 
\left (
\begin{array}{ccc}
m_{eff,0}^{11} ~I_e & m_{eff,0}^{12} 
~\sqrt{I_\mu}  ~\sqrt{I_e} 
& m_{eff,0}^{13} ~\sqrt{I_e}  ~\sqrt{I_\tau} \\
 & & \\
m_{eff,0}^{21} ~\sqrt{I_\mu}  ~\sqrt{I_e} 
& m_{eff,0}^{22} ~I_\mu & m_{eff,0}^{23} 
~\sqrt{I_\mu}  ~\sqrt{I_\tau} \\
 & & \\
m_{eff,0}^{31} ~\sqrt{I_e}  ~\sqrt{I_\tau}
 & m_{eff,0}^{32} 
~\sqrt{I_\mu}  ~\sqrt{I_\tau} & 
m_{eff,0}^{33}  ~I_\tau
\end{array}
\right) \nonumber  \\
 = 
\left (
\begin{array}{ccc}
\sqrt{I_e} & 0 & 0 \\
0 & \sqrt{I_\mu} & 0 \\
0 & 0 & \sqrt{I_\tau}
\end{array}
\right)
\cdot 
\left (
\begin{array}{ccc}
m_{eff,0}^{11} & m_{eff,0}^{12} & m_{eff,0}^{13} \\
 & & \\
m_{eff,0}^{21} & m_{eff,0}^{22} & m_{eff,0}^{23} \\
 & & \\
m_{eff,0}^{31} & m_{eff,0}^{32} & m_{eff,0}^{33} 
\end{array}
\right)
\cdot 
\left (
\begin{array}{ccc}
\sqrt{I_e} & 0 & 0 \\
0 & \sqrt{I_\mu} & 0 \\
0 & 0 & \sqrt{I_\tau}
\end{array}
\right) 
\label{factor}
\eea 
at $m_{SUSY}$. We evaluate subsequently
the integrals $I_{e,\mu,\tau}$ appearing in this
renormalization.
However, we can already extract some important 
qualitative information from (\ref{factor}).

$\bullet$
We first note that because of the factorization in
(\ref{factor}), although the individual masses and 
mixings get modified, any mass matrix which is
singular with a vanishing determinant -
leading to a zero mass eigenvalue -
remains so at the one-loop level. However,
one should expect modifications at the
two-loop level, which might be an interesting
mechanism for generating a non-trivial but large
neutrino-mass hierarchy.

$\bullet$
The Yukawa renormalization factors $I_i$ are less
than unity, and lead to the mass ordering
$m_{\nu_e} > m_{\nu_\mu} > m_{\nu_\tau}$, to the
extent that such naive flavour identifications are possible.

$\bullet$
One would expect that for values of $I_{\tau}$ 
substantially different from unity - which occur for
large
$\tan\beta$ in particular~\footnote{Most 
flavour-symmetry models in the literature assume
large $\tan\beta$.} -
the renormalization effects on the (23) sector would be especially
significant. However, there can be important effects even in the
first-generation sector. These can be significant for two reasons.
One is that, in view of the neutrinoless double-$\beta$ decay analysis
given above, very small mass differences may be
required for addressing the solar neutrino problem, so
we should keep even small renormalization effects in mind.
The other is that, when off-diagonal entries in
$m_{eff,0}^{ij}$ are large as in the sample texture
(\ref{GGtexture}), the $I_{\tau}$ renormalization effects
feed through into all differences in mass eigenvalues.

$\bullet$
In the case of the mixing angles, we 
recall that renormalization effects may either
enhance or suppress the mixing. In particular, it has been noted
in the case that $m_{eff,0}^{22} = m_{eff,0}^{33}$
and atmospheric-neutrino mixing is
maximal somewhere above the electroweak scale,
the maximal mixing may not survive down to
low energies if the $\tau$ Yukawa coupling is large,
at least for certain textures, depending on the magnitude of
the (23) entries. To illustrate this, we 
will make some generic
comments on the case of $2 \times 2$ 
mixing, and  we will 
return to neutrino-mixing effects  for the texture 
(\ref{GGtexture}) in the next section.

In the case of simple $ 2\times 2$ mixing,
we see from
\begin{equation}
\sin^2 2\theta_{23}  = 
\frac{4 
(m_{eff,0}^{23}) ^2}{( m_{eff,0}^{33} - 
m_{eff,0}^{22})^2 + 4 (m_{eff,0}^{23})^2}
\end{equation}
that the degeneracy between
$m_{eff}^{22}$ and $m_{eff}^{33}$ 
becomes important only if
$m_{eff}^{23}$ is of the same order as 
$m_{eff}^{33} - m_{eff}^{22}$.
It is known that large neutrino-mass hierarchies can be generated by
two-generation textures of the following forms
\cite{recent,ELLN2}
in the basis where the charged leptons are diagonal:
\bea
\left (
\begin{array}{cc}
x^2 & x \\
x & 1 
\end{array}
\right ),  ~
\left (
\begin{array}{cc}
1 & \pm 1 \\
\pm 1 & 1 
\end{array}
\right ), ~
\left (
\begin{array}{cc}
1 & 1 \\
-1 & -1 
\end{array}
\right ), ~
\left (
\begin{array}{cc}
1 & x \\
x & 1 
\end{array}
\right )
\label{tex1}
\eea
where the first solution has a large but
non-maximal mixing  in contrast with the others.
For the first texture \cite{Grah}, where 
$m_{eff,0}^{22} < m_{eff,0}^{33}$,
the renormalization-group effects on the mixing 
clearly are negligible, and 
the same is also true for the third texture, due to
the signs of the entries~\footnote{The 
renormalization of three-generation textures is more complicated, as
we see in the next section.}. In the case of
the second texture, one expects mild changes
despite the fact that $m_{eff,0}^{22} 
=  m_{eff,0}^{33}$, because $m_{eff,0}^{23} $ is large.
In the fourth texture \cite{degen1},
$m_{eff,0}^{22} 
=  m_{eff,0}^{33}$, and $m_{eff,0}^{23}$
is small, which is exactly the type of solution that
is very unstable under renormalization-group running.
Finally, we note that solutions with small hierarchies and large mixing
\cite{degen2}
of the type
\bea
\left (
\begin{array}{cc}
x & 1 \\
1 & x' 
\end{array}
\right )
\label{tex2}
\eea
are also expected to be stable under the renormalization group.

To complete this algebraic discussion of renormalization-group
effects, we now comment on the second case of interest,
in which the neutrino-mass texture is assumed to be
generated by non-renormalizable interactions
at some relatively low mass scale $\Lambda \sim 10^5$~GeV,
such as $\nu_L \nu_L H H / \Lambda$.
In this case, there are no neutrino Dirac couplings to be
renormalized, so the renormalization-group running between 
$M_{GUT}$ and $\Lambda$ is the same as equations
(\ref{MASSES}) to (\ref{factor}), with
Yukawa couplings only for quarks and charged leptons.
Below the scale $\Lambda$, $m_{eff}$ runs
in the same way as we discussed previously below $M_N$,
but the range of scales over which the
renormalization must be computed is greatly reduced.

\section{Numerical Results}

We now present some numbers for $I_\tau$ and $I_\mu$,
in order to exemplify renormalization effects on the textures.
We take as illustrative initial conditions~\footnote{Although the runnings
of the gauge couplings and of
$h_t$ factor out, they nevertheless affect the magnitude of $h_\tau$
and hence the exact value of $I_\tau$ that one derives.}
$\alpha_{GUT}^{-1} = 25.64$, 
$M_{GUT} =  1.1 \cdot 10^{16}$~GeV and
$m_{SUSY} = 1$ TeV.
We also take $h_t = 3.0$, and choose $h_b / h_\tau$ 
such that an intermediate scale $M_N$ is
consistent with the observed pattern of fermion masses
\footnote{The choice of input parameters needed to
reproduce exactly the observed fermion masses depends on $\tan\beta$,
but incorporating this refinement is unnecessary for our purposes.
We comment only that, for small
$\tan\beta$, an intermediate scale $M_N$ may be
consistent with the values of $m_b$ and $m_\tau$ measured at
low energies, at the cost of a certain 
deviation from bottom--tau mass unification
\cite{VB}, which may be $\sim 10 \%$
for $M_N \approx 10^{13}~{\rm GeV}$. 
However, this may be corrected~\cite{LLR}, if there is
sufficient mixing in the charged-lepton sector. For completeness,
we note that we use $h_N = 3.0$: this choice has a small impact
on the initial conditions at the scale $M_N$.}.
We use the physical $\mu$ and $e$ masses to fix $h_\mu, h_e$.
Finally, we take $M_N = 10^{13}$~GeV as our default,
mentioning later the effects with a different choice.
The values of $I_\tau$ and $I_\mu$ that we find with these inputs
are given in the first three columns of Table~1 and 
plotted in Fig.~1. These results may be used to
estimate the effects on the 
neutrino eigenvalues, mixings
and mass differences in the specific texture (\ref{GGtexture}),
as shown in the last three columns of Table~1 and in Fig.~2.

\begin{table}[tbp]
\begin{center}
\begin{small}
\begin{tabular}{|c|c|c|c|c|c|} \hline \hline
$h_\tau$  & 
$I_\tau$ & 
$I_\mu$  &
$m_{3}$  &
$m_{2}$ &
$m_{1}$  \\
\hline
         3.0 &   0.826  & 0.9955  & 0.866 & -0.952 &  0.997
\\ \hline
         1.2 &  0.873  & 0.9981 & 0.903 & -0.966 &  0.998 
\\ \hline
         0.48 &  0.9497   & 0.9994  & 0.962 & -0.987 & 0.9996
\\ \hline
        0.10 &  0.997   & 0.99997  & 0.9478 & -0.9993 & 0.99998 
\\ \hline
        0.013 &  0.99997   & 1.00000  & 0.99998 & -0.99999 & 1.00000
\\ \hline \hline 
\end{tabular}
\end{small}
\end{center}
\caption{
{\it  Values of $I_\tau$ and $I_\mu$, for $M_{N} = 10^{13}$~GeV
and different choices of $h_\tau$. Also tabulated are the three
renormalized mass eigenvalues calculated from the sample
texture (\ref{GGtexture}). 
}}
\end{table}

We see that the 
renormalization-group effects on the neutrino-mass
eigenvalues  are significant. Since they
are larger for the second- and third-generation leptons, 
as already commented,
we find the following ordering of the light neutrino
masses for textures with degenerate
eigenvalues at $M_N$: $m_{\nu_{e}} > 
m_{\nu_{\mu}} > m_{\nu_{\tau}}$,
which is {\it incompatible} with MSW solutions to the
solar-neutrino problem.
It is apparent from Table~1 and Fig.~2 that the breaking of
the neutrino-mass degeneracy in this model is unacceptable
for any value of $h_\tau$ corresponding to $1 < {\rm tan} \beta < 58$.

\begin{figure}[tbp]
\vspace*{-4.4 cm}
\centerline{\epsfig{figure=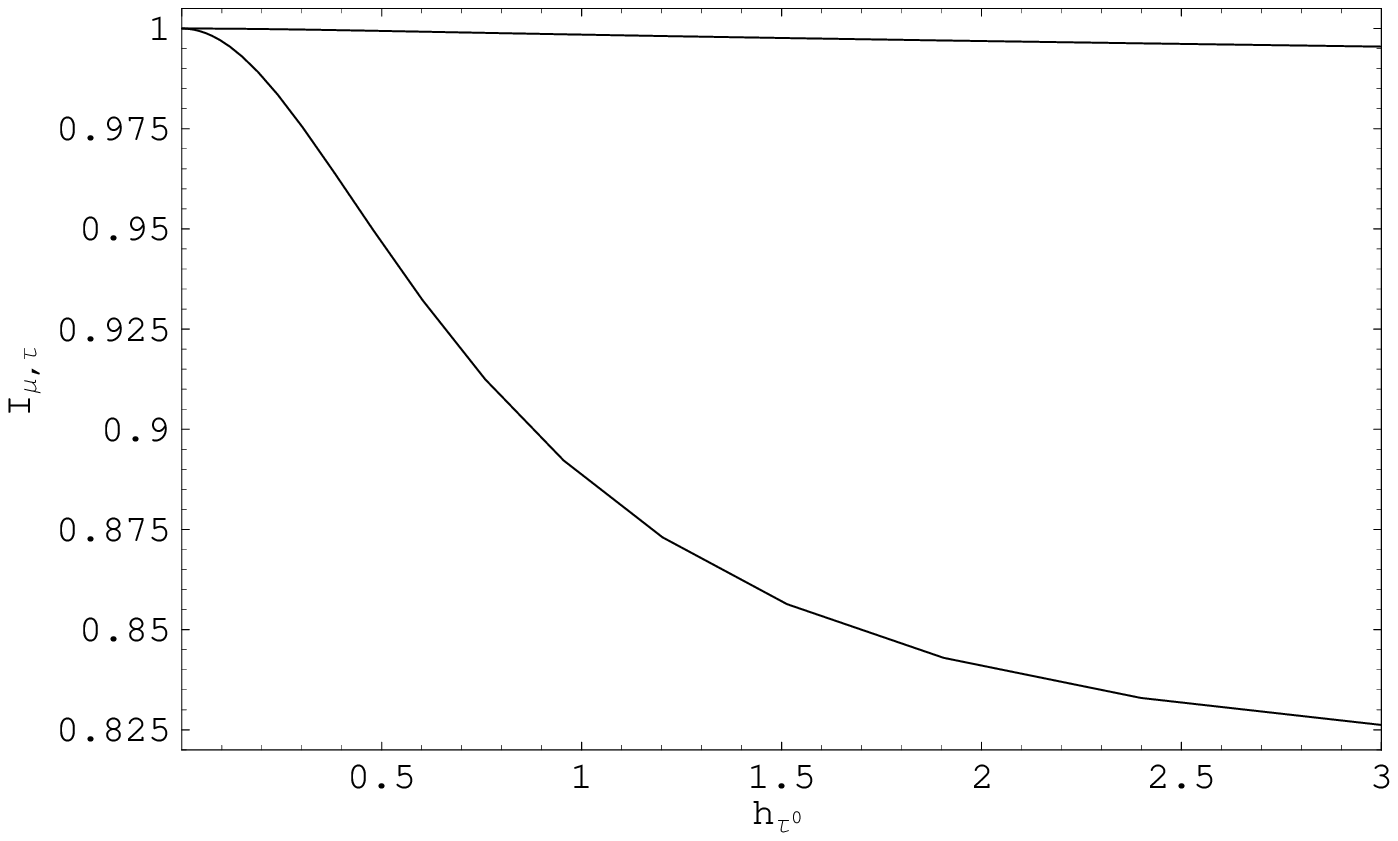,width=1.2\textwidth,clip=}}
\vspace*{-13.0 cm}
\caption{{\it Numerical values of
$I_{\mu}$ and $I_\tau$ for different 
initial values of $h_\tau$, assuming $M_N = 10^{13}$~GeV.
The values of $h_\tau^0 = $ 0.013, 0.03,
0.05 and 0.5 to 3  in steps
of 0.5 correspond to tan$\beta$
1, 3.8, 6.5, 43.8, 53.6, 56.3, 57.4, 57.9 and 58.2, respectively. 
}}
\vspace*{-2.4 cm}
\centerline{\epsfig{figure=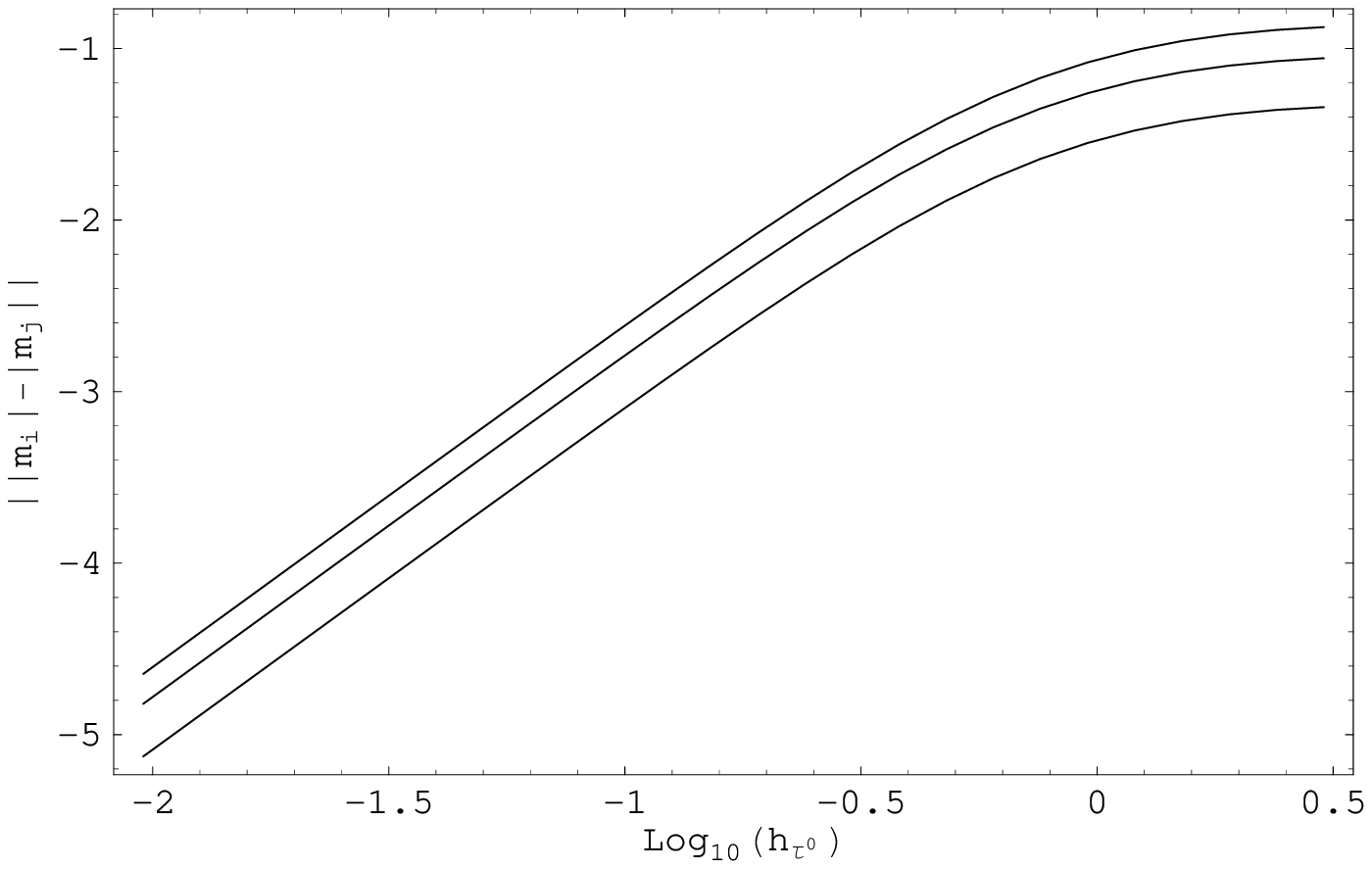,width=1.2\textwidth,clip=}}
\vspace*{-13.0 cm}
\caption{
{\it 
Renormalization of $m_{eff}$ eigenvalues for different 
initial values of $h_\tau$
corresponding to values of tan$\beta$ in the range 1 to 58,
assuming the particular
neutrino-mass texture (\ref{GGtexture}) and $M_N = 10^{13}$~GeV.
We see that the vacuum-oscillation scenario is never accommodated.}}
\end{figure}

We now discuss the renormalization of the neutrino 
mixing angles, using
as a particular example the texture (\ref{GGtexture}).
Initially we make a generic discussion, based
on analytic formulas, and then we illustrate the
discussion using two numeric examples.
We consider the following parametrization
of a perturbation from the initial
texture, motivated by the structure (\ref{factor}):
\bea
{m_{eff}'} \propto \,
\left (
\begin{array}{ccc}
0 &  {1\over\sqrt2} &  {1\over\sqrt2} 
(1+{\epsilon \over 2}) \\
& & \\
{1\over\sqrt2} &  {1\over2} &  -{1\over2} (1+{\epsilon \over 2})  \\
& & \\
{1\over\sqrt2} (1+{\epsilon \over 2}) &  -{1\over2}  
(1+{\epsilon \over 2}) &  {1\over2} (1+{\epsilon }) 
\end{array}
\right)
\label{GGtextureP}
\eea
where $\epsilon$ is a small quantity,
which might arise from renormalisation group running
or from some other higher-order effects
such as higher-dimensional non-renormalizable operators.
This perturbation lifts the degeneracy of the eigenvalues,
which are now given by
\bea
1, ~~ -1 - {\epsilon \over 4}, ~~1 + {3 \epsilon \over 4}
\nonumber 
\eea
To this order, the eigenvectors
are independent of $\epsilon$ and given by
\bea
V_1 = \left (
\begin{array}{c}
{1 \over \sqrt{3}} \\ \sqrt{ \frac{2}{3} } \\ 0 
\end{array}
\right ),~~
V_2 = \left (
\begin{array}{c}
{1 \over \sqrt{2}} \\ -{1 \over 2}  \\ -{1 \over 2} 
\end{array}
\right ), ~~ V_3 = 
\left (
\begin{array}{c}
{1 \over \sqrt{6}} \\ -{1 \over {2 \sqrt{3}}}  \\ {\sqrt{3}\over {2}} \\
\end{array}
\right )
\label{vec1}
\eea
so that the mixing expected in this
type of texture does not depend on $\epsilon$,
as long as it is non-zero.

However, this mixing is {\it not} bimaximal.
The vectors (\ref{vec1}) are also eigenvactors  of the
unrenormalised texture (\ref{GGtexture}).
Since this unperturbed texture has two exactly degenerate
eigenvalues, there is
arbitrariness in the choice of eigenvectors:
the vectors corresponding to the
two degenerate eigenvalues can be rotated to different
linear combinations, which are still
eigenvectors of the neutrino mass matrix and
still obey the orthogonality conditions.
One example is the choice
\bea
V_1 & = & \frac{1}{\sqrt{3}} V_1'  
+ \sqrt{\frac{2}{3}} V_3'  \nonumber \\
V_3 & = & \frac{1}{\sqrt{3}} V_3' 
-\sqrt{\frac{2}{3}} V_1'  \nonumber 
\eea
which gives
\bea
V_1' = \left (
\begin{array}{c}
0 \\ \frac{1}{\sqrt{2}} \\ -\frac{1}{\sqrt{2}} 
\end{array}
\right ),~~
V_2' = \left (
\begin{array}{c}
{1 \over \sqrt{2}} \\ -{1 \over 2}  \\ -{1 \over 2} 
\end{array}
\right ), ~~ 
V_3' = \left (
\begin{array}{c}
\frac{1}{\sqrt{2}} \\ \frac{1}{2} \\
\frac{1}{2} 
\end{array}
\right )
\label{vec2}
\eea
corresponding to bimaximal mixing: $\phi_1 = {\pi \over 4}$,
$\phi_2 = 0$ and $\phi_3 = {\pi \over 4}$. However, one cannot
in general expect this combination of eigenvectors to
be stable when the degenerate texture is perturbed,
and the above analysis shows that, indeed, it is not.
On the contrary,
it is the direction given by (\ref{vec1}) that is stable,
and the absence of the parameter $\epsilon$ in the eigenvectors
indicates that we may expect only minor modifications
in the mixing, for $\tau$ couplings in the range $3.0-0.013$.

We illustrate the instability of bimaximal mixing
and the stability of the eigenvectors (\ref{vec1}) with a
numerical analysis
of two extreme cases with $h_{\tau}^0 = 3$ and $ 0.013$.
Using the values of $I_{\tau ,\mu}$
given in Table 1, we determine the full
renormalized mass matrices to be:
\bea
m_{eff}^{1,ren} =
\left (
\begin{array}{ccc}
0 & 0.705 & 0.64 \\
0.705 & 0.497 & -0.45 \\
0.64 & -0.45 & 0.41
\end{array}
\right), ~~
m_{eff}^{2,ren} =
\left (
\begin{array}{ccc}
0 & 0.7071 & 0.7071 \\
0.7071 & 0.5 & -0.499992 \\
0.7071 & -0.499992 & 0.499985
\end{array}
\right), ~~
\eea
respectively, 
which are to be compared with the initial form 
(\ref{GGtexture}) of the texture.
Then, for 
\bea
U_1  =
\left (
\begin{array}{ccc}
0.5804 & 0.8143 & 0.0065 \\
-0.7075 & 0.5003 & 0.4992 \\
0.4032 & -0.2943 & 0.8665
\end{array}
\right), ~
U_2  =
\left (
\begin{array}{ccc}
0.57864 & 0.815578 & 0.00274 \\
-0.7071 & 0.5 & 0.5 \\
0.406418 & -0.29126 & 0.866021
\end{array}
\right)
\eea
we find that
\bea
m_{eff,diag }^{1,ren} = 
U_1 \cdot m_{eff}^{1,ren} \cdot U_1^T =
\left (
\begin{array}{ccc}
0.997 & 0 & 0 \\
0 & -0.952 & 0 \\
0 & 0 & 0.866
\end{array}
\right)
\eea
and 
\bea
m_{eff,diag }^{2,ren} = 
U_2 \cdot m_{eff}^{2,ren} \cdot U_2^T =
\left (
\begin{array}{ccc}
1.00000 & 0 & 0 \\
0 & -0.99999 & 0 \\
0 & 0 & 0.99998
\end{array}
\right)
\eea
reflecting the reverse mass ordering
mentioned above. 
On the other hand, as we have already remarked, the
eigenvectors (and thus the mixing matrices) are 
stable.
It is easy to check that
the matrices $U_{1,2}$ are unitary ones,
and in the notation of (\ref{param}), correspond 
to the values
\bea
\phi_1 \approx -0.327, ~~\phi_2 \approx 0.415, ~~
\phi_3 \approx  -0.884
\eea
In our notation, atmospheric-neutrino mixing 
is controlled by
the parameter $\phi_1$, for which
we find $\sin^2 2\phi_1 \approx 0.37$, whereas
solar-neutrino mixing is controlled by
$\phi_3$, for which we find $\sin^2 2\phi_3 \approx 0.96$.
We see therefore that even small
perturbations of exact neutrino
degeneracy cause large effects on the neutrino
mixing angles, which then {\em conflict with
the combined bounds from neutrinoless
double-$\beta$ decay and oscillation data}. This example shows
that the mixing
differs significantly from that postulated in
the unperturbed degenerate texture, an
effect not visible in a naive
$2 \times 2$ analysis.

Up to now we have been discussing the situation
where light neutrino masses arise through the see-saw
mechanism, and therefore $m_{eff}$ arises at
a scale $10^{13}$~GeV. However,
if $m_{eff}$ arises at a significantly
lower scale $\Lambda$, for example via effective operators of the form
$\nu_L \nu_L H H / \Lambda$, as discussed at the end of the previous
section,
the integrals $I_{\tau}$ and $I_{\mu}$ are now much closer to unity.
This happens because
(i) the range where $m_{eff}$ runs is significantly decreased,
and (ii) the starting value of $h_{\tau}$ at 
$\Lambda$ is also smaller, due to the 
run from $M_{GUT}$ to $\Lambda$ being over a 
relatively wide range. Shown in
Table~2 and Fig.~3 are our calculations of
$I_\tau$ and $I_\mu$, using the same parameters as in 
Table~1 and Fig.~1, except that now we run down from
$\Lambda = 10^{5}$~GeV, corresponding
to a scenario where $h_N$ is small
or zero~\footnote{Since the logarithmic
range of renormalization-group running is short in this
case, finite renormalization effects may be relatively more
significant than in the $M_N = 10^{13}$~GeV case. However,
their evaluation requires detailed modelling of thresholds,
which lies beyond the scope of this paper. We consider it unlikely
that the qualitative conclusions of this paper would be affected
by their inclusion.}.
The effects on the eigenvalues appear in Table~2 and 
Fig.~4, from which we see that, although the 
mass ratios are now closer to unity 
than in the previous case,
the effects of the running can still 
not be neglected,
when compared to the small mass differences
required by the solar neutrino data.
We again find that the full range $1 < {\rm tan} \beta < 58$
is excluded. On the other hand, the effect on the
mixing angle is similar to the previous case,
since it is practically unchanged under small perturbations,
as we discussed earlier.

\begin{table}[h]
\begin{center}
\begin{small}
\begin{tabular}{|c|c|c|c|c|c|} \hline \hline
$h_\tau$  & 
$I_\tau$ & 
$I_\mu$  &
$m_{3}$  &
$m_{2}$ &
$m_{1}$  \\
\hline
         3.0 &   0.973   & 0.9988  
& 0.9795 & -0.9929 &  0.9992
\\ \hline
         1.2 &  0.975   & 0.9995 
& 0.9815 & -0.9937 &  0.9996 
\\ \hline
         0.48 &  0.986   & 0.9998  
& 0.9897 & -0.9965 & 0.9999 
\\ \hline
        0.10 &  0.9999   & 0.99999  & 
0.9993 & -0.9997 & 0.99999 
\\ \hline
        0.013 &  0.99999   &  1.000000
& 0.999992 & -0.999997 & 1.00000
\\ \hline \hline 
\end{tabular}
\end{small}
\end{center}
\caption{
{\it  Values of $I_\tau$ and $I_\mu$, for $\Lambda = 10^{5}$~GeV
and different choices of $h_\tau$. Also tabulated are the three
renormalized mass eigenvalues calculated from the sample
texture (\ref{GGtexture}). }}
\end{table}

\begin{figure}[tbp]
\vspace*{-4.4 cm}
\centerline{\epsfig{figure=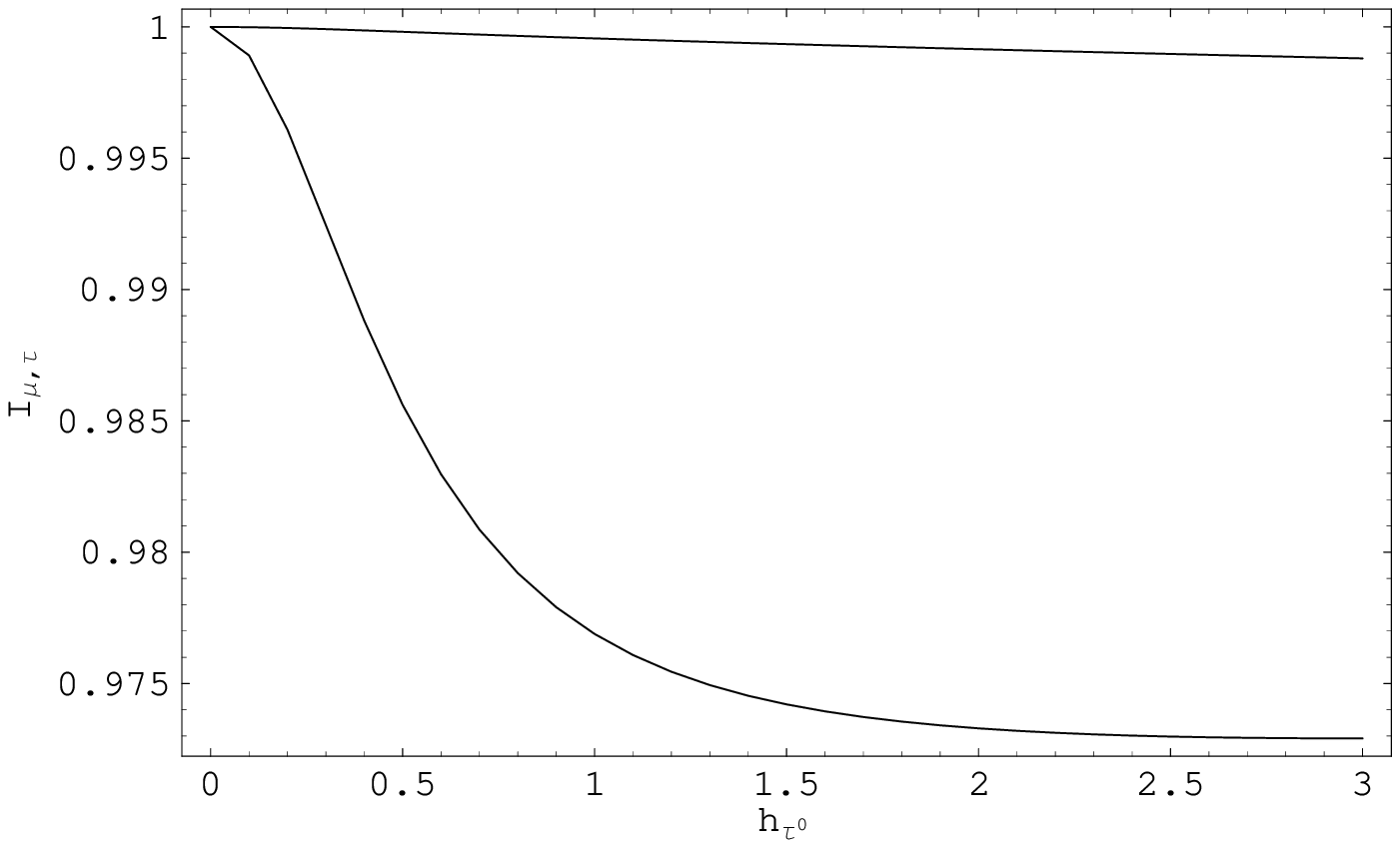,width=1.2\textwidth,clip=}}
\vspace*{-13.0 cm}
\caption{{\it Numerical values of
$I_{\mu}$ and $I_\tau$ for different 
initial values of $h_\tau$, assuming $\Lambda = 10^{5}$~GeV.
The corresponding values of tan$\beta$ are roughly 
the same as in the previous case.}}
\vspace*{-2.4 cm}
\centerline{\epsfig{figure=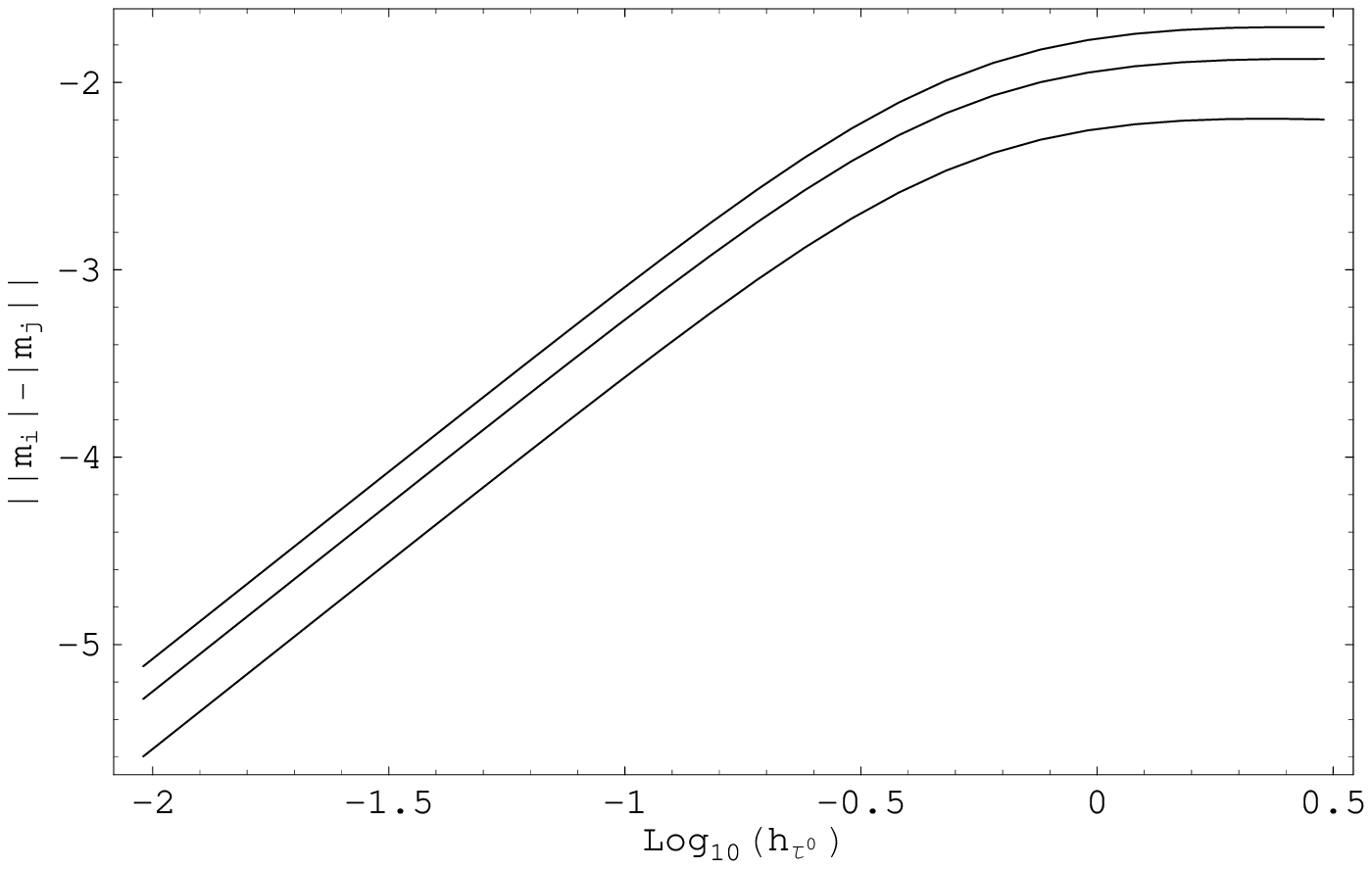,width=1.2\textwidth,clip=}}
\vspace*{-13.0 cm}
\caption{
{\it 
Renormalization of $m_{eff}$ eigenvalues for different 
initial values of $h_\tau$ 
corresponding to values of tan$\beta$ in the range 1 to 58,
assuming the particular
neutrino-mass texture (\ref{GGtexture}) and $\Lambda = 10^{5}$~GeV.
We see again that the vacuum-oscillation scenario is never accommodated.}}
\end{figure}

We comment finally on the 
renormalization-group effects in the 
minimal non-supersymmetric Standard Model.
The evolution equation for $m_{eff}$ is
\begin{equation}
16 \pi^2 {d m_{eff} \over d  t}
= ( - 3 g_2^2 + 2 \lambda + 2 S ) m_{eff}
- {1\over 2} ( (m_{eff} (Y_e^\dagger Y_e) + (Y_e^\dagger Y_e)^T m_{eff}),
\end{equation}
where $\lambda$ is the Higgs coupling:
$M_H^2 = \lambda v^2$,
and $ S \equiv Tr( 3 Y_u^\dagger Y_u + 3 Y_d^\dagger Y_d + Y_e^\dagger Y_e
) $~\cite{Bab}. Although the running of
$m_{eff}$ differs from the MSSM, the structure is similar.
In particular, the contributions proportional to
$g_2$, $\lambda$ and $S$ are the same for all entries,
and thus the exponential factors that are
obtained by integrating the renormalization-group
equations multiply all entries,
just as $I_t$ and $I_g$ did in the case of the MSSM.
The term that 
affects the relative runnings of the various entries
is again 
$ {1\over 2} ( (m_{eff} (Y_e^\dagger Y_e) + (Y_e^\dagger Y_e)^T m_{eff}$,
though with a sign opposite from the MSSM, meaning that now
the Yukawa couplings increase the entries
in $m_{eff}$. An important feature
in the Standard Model,
since it has only one Higgs field, is that the mass
hierarchies between fermions with opposite electroweak hypercharge
have to arise purely from the
Yukawa couplings. Hence the 
starting value of the $\tau$ coupling
is small in this case,
and therefore the effects are
expected quantitatively
to be similar to those in the 
low-$\tan\beta$ MSSM, 
but in the opposite direction.
This is interesting, since whereas
starting from degenerate-mass neutrinos
in the MSSM we expect low-energy neutrino hierarchies
of the type
$m_{\nu_\tau} < m_{\nu_\mu} < m_{\nu_e} $,
in the Standard Model we expect the opposite ordering
of masses: $m_{\nu_\tau} > m_{\nu_\mu} > m_{\nu_e} $,
which is the right sign for MSW solutions of the
solar-neutrino problem. However, the SM case shares with
the MSSM case the instability in the bimaximal mixing.
This means that the renormalized mass matrix is
again incompatible with the combined constraints from neutrinoless
double-$\beta$ decay and oscillation data, even 
though the breaking of the
mass degeneracy might appear compatible with the
MSW solution to the solar-neutrino problem.

\section{Conclusions}

We have studied in this paper the circumstances under which
neutrino masses can be degenerate and close to the
present upper bounds from Tritium $\beta$ decay and
astrophysics. We find that such schemes are 
severely constrained. In particular,
the new upper limit on neutrinoless double-$\beta$ decay~\cite{KKG},
in combination with the rest of the solar-neutrino
data, seems to exclude even
the large-angle MSW solution to the solar-neutrino problem,
and thus degenerate neutrinos may be compatible only
with  vacuum oscillations.
However, in this case extreme mass degeneracy to one part in $10^{10}$
is required. 

Even if such a degeneracy is guaranteed
by a symmetry at a certain scale, we find that renormalization
group effects lift this degeneracy.
In the MSSM with light neutrino masses arising through the
see-saw mechanism, the effects on the eigenvalues
are larger for larger $\tan\beta$, and have the wrong sign
for MSW solutions.
For a given $\tan\beta$, the effects are
reduced if $m_{eff}$ arises
via a non-renormalizable operator
such as $\nu_L \nu_L H H / \Lambda$  at a 
significantly lower scale 
than the $10^{13}$~GeV required by the see-saw.
Even in this case, however, the effects may not be neglected,
in view of the extreme mass degeneracy that is required.
Moreover, we find that even small perturbations shift the 
neutrino mixing angles by finite amounts, violating
the combined constraints from neutrinoless double-$\beta$ decay
and oscillation data.
Finally,  we find in the  minimal Standard Model 
renormalization effects that 
are qualitatively similar to those of the low-$\tan\beta$ MSSM,
but with opposite signs, thus leading to 
reversed low-energy neutrino-mass ordering. In this case,
the large-angle MSW solution may survive, but the instability
in the degenerate neutrino mixing angles means that the
combined constraints from
neutrinoless double-$\beta$ decay and oscillation
data are still violated.

Our analysis indicates that degenerate neutrino-mass textures
have many problems when renormalization effects are taken into
account. These results may provide hints on the appropriate
framework for flavour symmetries, with Abelian models~\cite{abel,LLR,Grah}
apparently favoured.

\vspace*{0.35 cm}
\noindent
{\bf Acknowledgements:} We thank A. Casas, J.R. Espinosa, 
Y. Nir and S. Pakvasa, for comments on the paper.

\end{document}